\documentclass[sigconf]{acmart}

\usepackage{dirtytalk}
\usepackage{makecell}
\usepackage{indentfirst}
\usepackage{graphicx}
\usepackage{subcaption}
\usepackage{caption}
\usepackage{booktabs}
\usepackage{multirow}
\usepackage{soul}  
\usepackage{siunitx}
\usepackage{xkeyval}
\usepackage{longtable}
\usepackage{array}
\AtBeginDocument{%
  }

\setcopyright{acmlicensed}
\copyrightyear{2018}
\acmYear{2018}
\acmDOI{XXXXXXX.XXXXXXX}

\acmConference[Conference acronym 'XX]{Make sure to enter the correct
  conference title from your rights confirmation emai}{June 03--05,
  2018}{Woodstock, NY}
\acmISBN{978-1-4503-XXXX-X/18/06}

\newcommand{\todo}[1]{\textcolor{black}{\textbf{[#1]}}}

\copyrightyear{2026}
\acmYear{2026}
\acmConference[CHI '26]{Proceedings of the 2026 CHI Conference on Human Factors in Computing Systems}{April 13--17, 2026}{Barcelona, Spain}
\acmBooktitle{Proceedings of the 2026 CHI Conference on Human Factors in Computing Systems (CHI '26), April 13--17, 2026, Barcelona, Spain}
\thanks{Accepted to CHI 2026. This is the authors' preprint version.}

\begin{document}

\ifdefined\reviewMode

    \newcommand{\rev}[1]{\textcolor{red}{#1}} 
    \renewcommand{\todo}[1]{\textcolor{red}{\textbf{[TODO: #1]}}}

\else

    \newcommand{\rev}[1]{#1} 
\fi

\author{Irene Hou}
\affiliation{%
  \institution{UC San Diego}
  \city{La Jolla, CA}
  \country{USA}}
\email{ihou@ucsd.edu}
\orcid{0009-0008-0511-7685}

\author{Zeyu Xiong}
\affiliation{%
  \institution{ETH Zurich}
  \city{Zurich}
  \country{Switzerland}}
\email{zeyu.xiong@inf.ethz.ch}
\orcid{0000-0002-3652-1890}

\author{Philip J. Guo}
\affiliation{%
  \institution{UC San Diego}
  \city{La Jolla, CA}
  \country{USA}}
\email{pg@ucsd.edu}
\orcid{0000-0002-4579-5754}

\author{April Yi Wang}
\affiliation{%
  \institution{ETH Zurich}
  \city{Zurich}
  \country{Switzerland}}
\email{april.wang@inf.ethz.ch}
\orcid{0000-0001-8724-4662}

\renewcommand{\shortauthors}{Hou et al.}
\begin{abstract}

Instructors are increasingly experimenting with AI chatbots for classroom support. To investigate how instructors adapt chatbots to their own contexts, we first analyzed existing resources that provide prompts for educational purposes. We identified ten common categories of customization, such as persona, guardrails, and personalization. We then conducted interviews with ten university STEM instructors and asked them to card-sort the categories into priorities. We found that instructors consistently prioritized the ability to customize chatbot behavior to align with course materials and pedagogical strategies and de-prioritized customizing persona/tone. However, their prioritization of other categories varied significantly by course size, discipline, and teaching style, even across courses taught by the same individual, highlighting that no single design can meet all contexts. These findings suggest that modular AI chatbots may provide a promising path forward. We offer design implications for educational developers building the next generation of customizable classroom AI systems.
 
\end{abstract}

\newcommand{\sys}{SYSNAME}
\title{``Bespoke Bots'': Diverse Instructor Needs for Customizing Generative AI Classroom Chatbots}

\ccsdesc[500]{Human-centered computing~Empirical studies in HCI}

\keywords{Human-centered AI, generative AI, educational chatbots, instructors, customization, participatory design}


\maketitle

\section{Introduction}

Instructors are increasingly experimenting with AI chatbots as classroom assistants. 
Across higher education, generative AI chatbots have been used to answer student questions, provide formative feedback, generate practice and assessment materials, and even integrate with learning management systems~\cite{chang2023educational, macneil2023experiences, google2025gemini, martin2024systematic, kazemitabaar2024codeaid}. 
Such deployments suggest that chatbots can help reduce bottlenecks in large courses and serve as an intermediary step before students seek human help~\cite{liffiton2023codehelp}. 
However, instructors' needs are heterogeneous. 
Instructors value tailoring their courses~\cite{vermette2019freedom, vermette2022uncovering} because what works in one setting may fail in another; pedagogy is inherently shaped by course format and size, institutional norms, and instructor preferences. 
This extends to how instructors desire to adapt chatbots to their classrooms~\cite{lau2023ban, hedderich2024a}. 
It also makes it difficult for any single ``course bot'' to fit all contexts.

Although approaches like prompt engineering~\cite{arawjo2024chainforge} or retrieval-augmented generation~\cite{lewis2020retrieval, maryamah2024chatbots} allow the assembly of custom chatbots, there remain challenges in adapting these approaches to classroom use. 
Instructional context requires chatbots to balance multiple goals simultaneously, such as accurate course content, pedagogical strategies, tone, and guardrails.
This can easily become  entangled in a single prompt, leading to unpredictable behavior such as switching personas mid-conversation~\cite{koyuturk2023developing}.
At the same time, classrooms demand reproducibility and shareability, yet evaluating and standardizing chatbot behaviors remains difficult~\cite{zamfirescu2023conversation, kim2024evallm},  making it hard to reuse or adapt designs across courses. 

Recent work on agentic AI workflows points to a promising alternative~\cite{du2024survey, sapkota2025ai}. 
Instead of relying on a monolithic prompt, instructors could compose a roster of modular agents, what we call \textit{teachable teammates}, each bound to distinct roles, syllabus rules, rubrics, and data boundaries (e.g. a grader bot bound to rubrics, a lab coach bot with safety protocols). This modularity could provide both flexibility and control.

However, before building such systems, we must first understand how instructors themselves would decompose their needs into such roles and priorities. \rev{Although modular and agentic workflows are becoming technically feasible, it remains unclear how instructional needs align with or challenge modular boundaries. Understanding where instructors’ needs converge versus diverge clarifies which customization capabilities could be standardized across contexts and which must remain flexible and instructor-driven, providing insight necessary to inform future design, whether modular, agentic, or otherwise.} Without this knowledge, efforts to design reusable chatbot systems risk missing the realities of day-to-day teaching.

\begin{itemize}
    \item \textbf{RQ1: What are instructors’ goals, challenges, and priorities for customizable classroom chatbots?}
    \item \textbf{RQ2: How do instructors’ customization needs converge and diverge across different teaching contexts (e.g., class size, course type, institution)?}
\end{itemize}

To answer these questions, we first conducted a content analysis on 182 prompts and features from public resources to identify an initial need space for customizing classroom chatbots.
We then conducted 10 semi-structured interviews with university STEM (Science, Technology, Engineering, and Mathematics) instructors who had prior experience using AI in their courses, paired with a card-sorting activity for feature prioritization. By empirically surfacing how instructors decompose and prioritize chatbot customization, we ground and extend technical visions of modular and agentic workflows. Building on this framing, our study contributes: 1) an empirical account of how instructors envision customizing classroom AI chatbots, including opportunities and obstacles, 
2) a synthesis of existing needs and their prioritization by instructors from the card-sorting activity,
3) implications for the design of future modular, agentic workflows in classroom chatbots. 

\section{Related Works}

\subsection{Instructors and the Customization of Classroom Technologies}
Instructors often customize their courses to fit pedagogy and student needs. Prior research highlights how teachers adapt diverse instructional strategies and technologies across contexts~\cite{bergmann2012flip, martin2020examining, messer2024automated, gorder2008study}. 
Such variation can be influenced by factors like instructors' computer self-efficacy~\cite{ertmer2010teacher}, institutional support~\cite{zhao2002conditions}, and time constraints~\cite{vermette2019freedom}. 
Course formats also vary: lab-based courses on collaboration~\cite{shibley2002influence, maughan2025integrating}, online courses on delivery mechanisms~\cite{chen2019improv}, and large courses on institutional infrastructures~\cite{martin2020examining}. Vermette et al. found that K–12 teachers actively build ``digital classroom ecosystems'' by combining and personalizing productivity suites, LMS, and plugins~\cite{vermette2019freedom}. In addition, they observed that customization among instructors is not just an individual activity---instructors often share ideas and troubleshoot issues through ``communities of practice''~\cite{wenger1999communities, hur2009teacher, vermette2019freedom}. Field deployment of a customization-sharing platform showed that instructors valued exchanging and adopting each other’s LMS modifications, which reduced their experimentation costs and improved awareness of features~\cite{vermette2022uncovering}.

While prior work demonstrates the value of customizing static or one-way tools such as LMS~\cite{instructure2021directshare, instructure2021canvascommons, vermette2022uncovering}, these contexts differ from emergent chatbot systems. With static tools, instructors can preview and control content before it is delivered. In contrast, chatbots are nondeterministic, generating responses that shift with student queries and cannot be fully anticipated~\cite{koyuturk2023developing}. This uncertainty raises new concerns---whether tone is appropriate, explanations scaffold learning, or accuracy drifts over time~\cite{hedderich2024a}. As a result, instructors may seek different forms of customization, such as guardrails, role assignment, or mechanisms to test and refine outputs. Our study examines this new context of customizing interactive AI systems.

\subsection{Generative AI and Chatbots in Education}

Consumer-available large language models (LLMs) such as ChatGPT~\cite{openai2025chatgpt}, Claude~\cite{anthropic2025claude}, and Gemini~\cite{google2025gemini} have introduced new opportunities and challenges for education. 
Early work documented instructors’ and students’ initial encounters, both positive and negative, with AI in the classroom~\cite{labadze2023role}, particularly in computing education~\cite{lau2023ban, hou2024effects, prather2024the}. In a review of 67 studies on the role of chatbots in education, benefits for students included homework support, personalized learning, and skill development~\cite{labadze2023role}. 
For educators, benefits included time-saving assistance and improved pedagogy. Although some instructors remain reluctant to adopt AI tools in their classroom, citing fears of over-reliance, ethical concerns, and inequities, others have acknowledged \textit{``resistance is futile''}~\cite{lau2023ban}. 

With student use of AI chatbots becoming widespread~\cite{hou2025evolving, stohr2024perceptions}, researchers have started exploring how instructors adapt by developing custom AI tools and AI-integrated classrooms~\cite{vadaparty2024cs1, denny2024desirable, atmosukarto2021enhancing, kapoor2025guardrails, chen2023artificial}. Tools have been introduced to make chatbot customization more accessible for educators. For instance, ChatbotBuilder allows instructors to manage dialogue flow in cyberbullying education~\cite{hedderich2024a}. Interest in ``bespoke'' educational chatbot systems has grown, and early deployments of bespoke chatbots have seen benefits for students and educators~\cite{hedderich2024a, liffiton2023codehelp}. Kazemitabaar et al's~\cite{kazemitabaar2024codeaid} CodeAid, an AI assistant that explains code but avoids giving direct answers, deepened students' conceptual understanding. Particular attention has been placed on designing pedagogical chatbots that provide scaffolding --- balancing AI assistance with guidance and the right amount of friction~\cite{denny2024desirable}. Other bespoke chatbots like GPTeach have been deployed to help instructors practice pedagogical skills~\cite{markel2023gpteach}. 

Although prior work has examined how AI assistants could support pedagogical interventions~\cite{liffiton2023codehelp}, assist instructors in course-related tasks~\cite{lau2023ban}, and improve teaching~\cite{markel2023gpteach}, there remain gaps in our understanding of the design space across instructional contexts. Our work seeks to identify how instructors actively desire the tailoring of such systems to align with their needs, which can differ across class sizes, disciplines, and teaching styles. We aim to more broadly capture instructors’ design priorities and desired controls.

\section{Content Analysis of Existing Prompts and Resources}
\label{sec:feature}
\label{sec:solution-space}

\begin{table*}[t]
\footnotesize
\caption{Ten feature categories for AI course chatbot customization used in the card-sorting activity as shown to participants. Categories are derived from content analysis of open-source prompts, as described in Sec. \ref{sec:feature}.}
\label{tab:chatbot-features}

\begin{tabular}{p{3cm} p{5cm} @{\hspace{0.5cm}} p{7cm}}

\toprule
\textbf{Category} & \textbf{Description} & \textbf{Examples of potential customizations} \\
\\
\midrule




\textbf{Pedagogical strategy} & Defines instructional method or pedagogical style that shapes interactions with students &
Socratic questioning; worked examples; adaptive questioning; scaffolding hints; peer learning; discovery learning \\
\midrule

\textbf{Course material} & Specifies the background course content and material the chatbot draws from &
Course/topic; lecture materials; textbooks; syllabus; assignments \\
\midrule

\textbf{Personalization} & Adapts responses to unique characteristics of individual learners &
Student skill level; student interests (e.g. sports, music); individual needs \\
\midrule

\textbf{Constraints/guardrails} & Imposes explicit guidelines, rules, and restrictions in chatbot behavior &
Response length limit; cite sources; no direct answers; flagging harmful interactions; restricting types of student queries \\
\midrule

\textbf{User environment} & Specifies the platforms, settings, or contexts in which the chatbot is deployed &
Canvas; Moodle; Discord; Piazza; office hours; lecture; lab; at home \\
\midrule

\textbf{Task/objective} & Defines the specific instructional goal or task the chatbot supports &
Explain concepts; generate questions; fact-check; provide example assessment problems \\
\midrule

\textbf{Course management} & Supports instructors in handling classroom logistics or administration &
Create student groups; provide grading assistance; collect course metadata; answer forum questions; summarize common questions \\
\midrule

\textbf{Persona/tone} & Defines the chatbot’s identity, role, and communicative tone &
Expert tutor persona; peer student persona; motivational tone; humorous tone; academic tone; neutral tone \\
\midrule

\textbf{Content format}& Specifies the desired structure or medium of the chatbot’s output &
Code snippets; diagrams; LaTeX/equations; tables; multi-media formats \\
\midrule

\textbf{Community sharing} & Enables reuse, adaptation, and distribution of chatbot configurations &
Shareable course templates; department templates; university templates; prompt libraries; remix options; adaptable course configurations \\

\midrule

\end{tabular}
\normalsize
\end{table*}

\subsection{Content Analysis Methods} We sought to identify common, concrete strategies that instructors use to customize chatbot behavior for their course. First, we conducted a qualitative content analysis~\cite{krippendorff2018content, elo2008qualitative} of prompts and chatbot features drawn from public educational resources directed at instructors (e.g., Microsoft Prompts for Education~\cite{microsoft2025promptsedu}, Anthropic Prompt Library~\cite{anthropic2025promptlibrary}, AI for Education~\cite{aiforeducation2025promptlibrary}, university prompt libraries~\cite{wharton2025promptlibrary, maastricht2025promptlibrary}, ChatGPT Study Mode~\cite{openai2025studymode}). Resources were selected via keyword search (``prompt library,'' ``AI for education,'' ``chatbot prompts for education'') and review of official or institutionally hosted repositories. \rev{We included prompt templates and feature descriptions (e.g., ChatGPT’s Study Mode) that allowed instructors to influence chatbot interactions and responses. Prompts were excluded  if they explicitly referred to contexts that were not related to educational or classroom settings (e.g. business report generators).} \rev{In total, our final corpus contained 182 prompts and feature descriptions.}

\rev{To guide our analysis, we used an inductive, bottom-up coding approach. We used inductive coding as there was no existing taxonomy that captured the range of educational chatbot prompt-based customizations. \rev{Text} prompts and feature descriptions were first independently coded by two team members. Descriptive codes initially captured the intended purpose of each prompt and feature description. For instance, the Anthropic ``Socratic Sage'' prompt~\cite{anthropic2025socraticsage} was coded with labels such as \textit{Socratic questioning} and \textit{critical thinking scaffold}. Across prompts, initial labels such as \textit{providing worked examples}, \textit{step-by-step hints}, or \textit{error analysis} frequently co-occurred, which were later consolidated into the higher-level thematic category \textbf{pedagogical strategy}. During this process, the team met periodically to discuss codes and reconcile coding disagreements until consensus was reached.}

\rev{\subsection{Content Analysis Results} Ten categories emerged through this process, representing common types of chatbot customization for instructional use (See Table~\ref{tab:chatbot-features} for a summary of each category and related examples).} \rev{Across our corpus, \textbf{task/objective} was the most frequently coded category, followed by \textbf{pedagogical strategy}, \textbf{content format}, \textbf{persona/tone}, \textbf{constraints/guardrails}, and \textbf{personalization.} \textbf{course material} and \textbf{course management} appeared infrequently in public prompt libraries such as Anthropic’s~\cite{anthropic2025promptlibrary}, but were frequently coded in resources explicitly marketed towards teachers, such as the AI for Education Prompt Library ~\cite{aiforeducation2025promptlibrary}. \textbf{User environment} customization was rare, appearing primarily as feature descriptions (e.g., integration into popular platforms like Google Drive). \textbf{Community sharing} appeared least frequently.} These ten categories informed the design of the card prototypes used in interviews (See Figure \ref{fig:cards}). Pilot interviews were conducted to further refine the category descriptions and interview protocol.

\section{Card-Sorting Interviews with Instructors}

To understand what STEM university instructors need from customized AI course chatbot systems, we conducted ten semi-structured interviews that incorporated a card-sorting feature prioritization activity. Interviews were conducted in English between June and August 2025, lasted 40–60 minutes, and participants received a \$20 gift card as compensation. This study was approved by our Institutional Review Board.

\subsection{Participants}

Participants were recruited via email through professional networks, institutional partners, and cold outreach. We sought university STEM instructors who had experimented with AI use in the classroom (see Table~\ref{tab:participants} for range of disciplinary backgrounds). They spanned North America (NA) and Europe (EU) universities, representing teaching contexts from small, project-based courses to massive lecture courses. Participants included university instructors and teaching assistants involved in course development and instruction, as well as industry-affiliated educators who teach university workshops (P9). We also interviewed an instructor (P2) responsible for developing a university-wide, custom chatbot initiative, during which he regularly consulted with other instructors. His perspective further triangulated our findings. This variation in participants allowed us to capture needs across many instructional formats and content areas. 

\begin{table*}[h]
\centering
\caption{Participant demographics and teaching contexts. Column `YoE' stands for Years of Experience. In the Teaching Focus column, CS stands for Computer Science, and HCI stands for Human-Computer Interaction.}
\label{tab:participants}
\resizebox{\textwidth}{!}{%
\begin{tabular}{llllllll}
\toprule
\textbf{ID} & \textbf{Role} & \textbf{Gender} & \textbf{Region} & \textbf{YoE} & \textbf{Teaching Focus} & \textbf{Class Size} & \textbf{Course Type} \\
\midrule
P1 & Instructor & F & EU & 5 & CS: Introductory Programming & 40 & Lecture-based \\
P2 & Developer/Instructor & M & EU & 20 & Physics & 30 & Lecture-based \\
P3 & Teaching Assistant & M & EU & 3 & HCI & 100-200 & Project and Lecture-based \\
P4 & Instructor & M & US & 5 & HCI & 100-200 & Project and Lecture-based \\
P5 & Instructor & M & US & 5 & Software Design & 30 & Capstone course \\
P6 & Instructor & M & US & 7 & Data Science & 100 & Project and Lecture-based \\
P7 & Teaching Assistant & M & EU & 11 & CS: Introductory Programming & 100-200 & Lecture-based \\
P8 & Instructor & F & US & 6 & CS: Data Structures & 100-200 & Project and Lecture-based \\
P9 & Instructor & F & US & 3 & Interaction Design & 120 & Project-based \\
P10 & Instructor & F & US & 6 & Math: Calculus & 4000 & Lecture-based and groupwork \\
\bottomrule
\end{tabular}
}
\end{table*}


\subsection{Interview Protocol and Analysis}
\begin{figure*}
  \centering
  \includegraphics[width=0.86\linewidth]{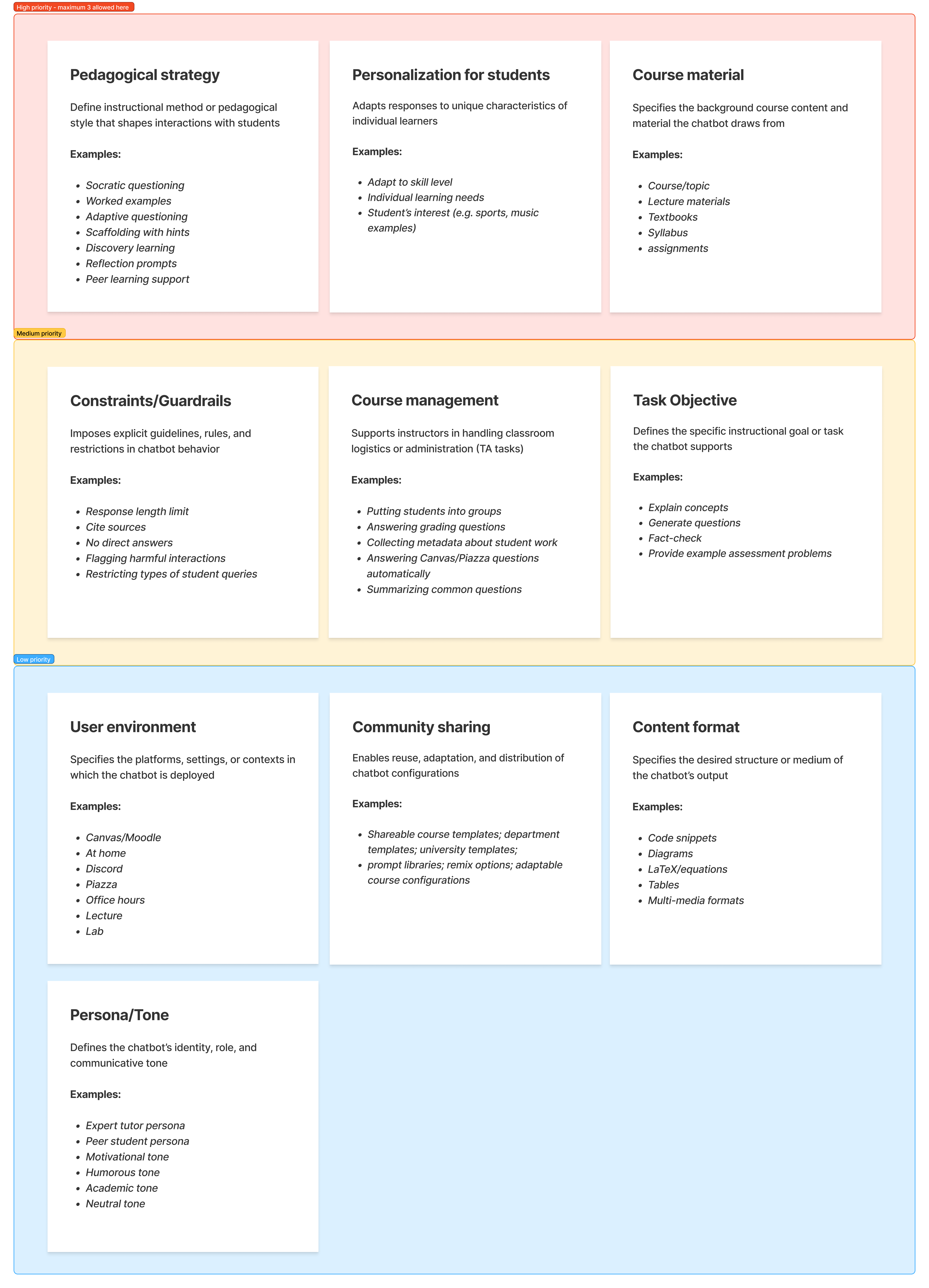}
  \caption{\rev{P6's finalized card-sorting interview artifact, with cards sorted into the ``high priority,'' ``medium priority,'' and ``low priority'' buckets. Card-sorting interviews were conducted via a collaborative online whiteboarding tool, with each card depicting a potential area of chatbot customization for instructors and example use cases. Refer to Table \ref{tab:chatbot-features} to see detailed card categories and examples.}}
  \label{fig:cards}
\end{figure*}

Participants first described their current teaching context and related challenges. Then, we invited them to imagine we were educational technology developers tasked with creating the `perfect' AI course chatbot for their classroom. This speculative framing, guided by methods from the field of speculative design~\cite{auger2013speculative, hoffman2022speculative}, encouraged participants to think aspirationally while remaining grounded in their actual teaching context. Next, drawing from the card-sorting HCI technique~\cite{spencer2009card}, we presented participants with ten cards via an online whiteboarding tool. Each card corresponds to a category from Table~\ref{tab:chatbot-features} and refers to a type of customization that can be applied to a chatbot \rev{(Sample card-sorting interview artifact shown in Figure~\ref{fig:cards})}. For example, customizing \textbf{pedagogical strategy} means applying specific instructional techniques to chatbot interactions.

\rev{To reduce anchoring bias, cards were shown to participants \textit{after} they had reflected on their current teaching context and challenges.} Participants were asked to think aloud while sorting each card into buckets labeled `high,' `medium,' or `low' priority (or `not important'), with only three cards allowed in `high.' Blank cards allowed the proposal of additional features. \rev{This card-sort method had three purposes: 1) contextualized instructors' priorities during think-aloud, 2) revealed instructors' underlying mental models, and 3) forced them to reason explicitly through trade-offs.} \rev{Discussion during the think-aloud activity helped establish shared understanding between participant and interviewer and minimized interpretive differences. Although each card included example use cases to clarify its category meaning, some instructors suggested additional constraints, formats, or interaction rules beyond those depicted on the cards. These elaborations surfaced naturally given the speculative framing and were included in our analysis.}

Interviews were consensually recorded and transcribed on Zoom. Three authors independently reviewed interview transcripts before meeting regularly to compare and discuss emerging patterns. We conducted a hybrid thematic analysis, starting with a deductive codebook derived from chatbot feature categories (Table ~\ref{tab:chatbot-features}), later inductively expanding it to capture new themes that emerged during analysis. \rev{We integrated the card-sorting data and interview transcripts by examining how instructors justified their rankings and trade-offs, allowing us to identify cross-cutting patterns in their reasoning and how customization needs were prioritized.}

\subsection{Study Scope and Limitations}

Our study is scoped to STEM university instructors' perspectives, with a skew towards computing education as many computing instructors are early adopters of classroom AI assistants~\cite{kazemitabaar2024codeaid, denny2024desirable, prather2025beyond}. Our findings may not apply to K-12, non-STEM, or informal learning environments. Additionally, while our card-sorting technique elicited rich, contrastive data, providing categories \rev{and examples} may have anchored instructor perspectives. We aimed to mitigate this by asking instructors to share challenges and experiences \textit{before} presenting cards, and also by providing blank cards to encourage broader perspectives. \rev{Although the card-sorting activity was grounded in discussion, and we did not observe substantive misinterpretations, a limitation of this method is that there may have been minor variations in how participants interpreted category boundaries, which may not be fully captured by interviewers.} Finally, while our participants had diverse backgrounds and were from EU/NA regions, reported perspectives are not globally-representative. These findings may not be generalizable across different educational cultures, experiences, or institutions.

\section{Results}

\begin{table*}[htbp]
\centering
\caption{Feature category importance rankings by individual participants. \textcolor{blue}{H} = High importance, M = Medium importance, L = Low importance, NI = Not important. The summary column shows counts in format (H/M/L/NI).}
\label{tab:feature-rankings-detailed}
\begin{tabular}{l|cccccccccc|cccc}
\toprule
\textbf{Feature Category} & \textbf{P1} & \textbf{P2} & \textbf{P3} & \textbf{P4} & \textbf{P5} & \textbf{P6} & \textbf{P7} & \textbf{P8} & \textbf{P9} & \textbf{P10} & \textbf{H} & \textbf{M} & \textbf{L} & \textbf{NI} \\
\midrule
Pedagogical strategy     & \textcolor{blue}{H} & L & L & \textcolor{blue}{H} & L & \textcolor{blue}{H} & \textcolor{blue}{H} & M & \textcolor{blue}{H} & \textcolor{blue}{H} & 6 & 1 & 3 & 0 \\
Task/objective          & \textcolor{blue}{H} & M & L & \textcolor{blue}{H} & L & M & M & L & L & M & 2 & 4 & 4 & 0 \\
Course material         & M & \textcolor{blue}{H} & \textcolor{blue}{H} & \textcolor{blue}{H} & \textcolor{blue}{H} & \textcolor{blue}{H} & M & M & \textcolor{blue}{H} & M & 6 & 4 & 0 & 0 \\
Constraints/guardrails  & M & L & \textcolor{blue}{H} & M & M & M & \textcolor{blue}{H} & M & M & M & 2 & 7 & 1 & 0 \\
User environment        & M & L & M & L & \textcolor{blue}{H} & L & M & \textcolor{blue}{H} & M & L & 2 & 4 & 4 & 0 \\
Course management       & L & NI & L & M & \textcolor{blue}{H} & M & NI & \textcolor{blue}{H} & L & \textcolor{blue}{H} & 3 & 2 & 3 & 2 \\
Community sharing       & L & \textcolor{blue}{H} & M & M & M & L & NI & \textcolor{blue}{H} & \textcolor{blue}{H} & M & 3 & 4 & 2 & 1 \\
Personalization         & L & L & \textcolor{blue}{H} & M & L & \textcolor{blue}{H} & M & L & M & \textcolor{blue}{H} & 3 & 3 & 4 & 0 \\
Content format          & M & NI & M & L & M & L & L & M & NI & M & 0 & 5 & 3 & 2 \\
Persona/tone            & \textcolor{blue}{H} & NI & M & L & L & L & L & L & L & M & 1 & 2 & 6 & 1 \\
\midrule
\multicolumn{15}{l}{\textit{Additional custom features suggested by participants (all ranked High):}} \\
\multicolumn{15}{l}{Cross-discipline relevancy (P1), easy set-up (P2), chatbot adaptability to student intent (P2),} \\
\multicolumn{15}{l}{cheap or low cost (P10), pedagogical training for TAs (P10)} \\
\bottomrule
\end{tabular}
\end{table*}

Below we elaborate on areas of commonality and divergence, organized around major feature categories, and how contextual factors shaped instructors’ needs. Refer to Table~\ref{tab:chatbot-features} for definitions of each feature category. Within our sample, we did not observe noticeable cultural differences across NA/EU instructors. Instructors' rankings of each feature category are shown in Table \ref{tab:feature-rankings-detailed}. 


\subsection{Where instructors converged on priorities}

Instructors consistently prioritized customizing chatbots around \textbf{course material}, \textbf{pedagogical strategy}, and \textbf{constraints/guard-rails}. On the opposite end, they \textit{de-prioritized} customizing chatbots for \textbf{content format} and \textbf{persona/tone}. 

\subsubsection{Instructors wanted custom chatbots to have access to \textbf{course-specific materials}.} The most consistent priority across instructors was the integration of course-specific materials into customized chatbot interactions. Six of ten participants ranked this as a top need, with the remainder placing it at medium. Instructors emphasized that without access to lecture slides, assignments, or textbooks, chatbots risk producing answers that are too generic or random to be reliable. P2, an instructor and developer of a customized chatbot for his university, observed that, \textit{``Lecture, discussion material, textbook, assignment materials --- that was what most of our instructors actually wanted from [developers], and chatbots strictly basing answers on the course materials provided, using the same nomenclature and definitions.''} He shared a story about \textit{``an instructor who used an unusual physics symbol for quantity was miffed that the [custom] bot used a symbol he didn't use in his course.''} \textbf{Course material} served as the baseline expectation for effective customization.

\subsubsection{Instructors wanted to customize chatbot \textbf{pedagogical strategy}.} Six of ten participants highly prioritized customizing the chatbot’s instructional style, such as scaffolding with hints, worked examples, or error analysis. Instructors wanted the chatbot to embody pedagogical methods that they used in class, rather than adopting a default ``neutral explainer'' voice. This feature was consistently framed as a way to align chatbot interactions with teaching philosophies. P9, who teaches interaction design workshops hosted in partnership with multiple universities, explained that it was challenging to ensure students were on the same page, especially when it came to peer learning, a pedagogical strategy she liked to employ. She prioritized \textbf{pedagogical strategy} because \textit{``there's benefit to having systemized structure with AI assistance so that students can say, `Oh, we're working on the same kind of prompts or examples.' If there's no consistency, students can't really collaborate.''} P8 hoped customizing for pedagogical strategies would best support students who were trying but struggling in the class, providing \textit{``some friction''} instead of direct answers like ChatGPT.

\subsubsection{Instructors found \textbf{constraints/guardrails} to be important but not the topmost priority.}

While only one instructor ranked \textbf{constraints/guardrails} as a top priority, seven placed them in the middle tier, suggesting they were viewed as a necessary baseline rather than a central customization goal. Instructors generally expected safety, accuracy, and guardrails to be ``built in,'' but did not see them as the most critical area to invest customization effort, compared to embedding course content or pedagogical strategies. \rev{One instructor broadened the \textbf{constraints/guardrails} category to include rules they wanted the chatbot to help enforce with students, highlighting that some instructors view guardrails as classroom practices rather than chatbot-only behaviors.} For instance, P5, who teaches an HCI course preparing students for industry, noted that he would want constraints that pushed students to write unit tests for code: \textit{``In industry code review sessions, they'll tell you, `Hey, we can't let this in without a unit test.' But my students are scared of missing deadlines. They'll let code through and say, `We'll test later.' I can't be their dad and check all the time if they're doing that, but I feel a chatbot might be able to.''}

\subsubsection{Instructors consistently de-prioritized \textbf{content format} and \textbf{persona/tone}.} 

Instructors consistently ranked both \textbf{content format} and chatbot \textbf{persona/tone} as low priorities for customization. For \textbf{content format}, no participant ranked it as high. While some noted that formatted outputs such as LaTeX (P10), tables (P5), or code blocks (P8) could be helpful, these were generally viewed as conveniences rather than core teaching needs. 

\textbf{Persona/tone} was viewed least favorably, with seven instructors explicitly placing it in the low importance or not important tier, often describing personality features as not useful or even potentially harmful. P2 felt strongly about \textit{\textbf{not}} allowing the chatbot to have any persona, instead desiring an academic or neutral tone. He expressed concern about personable AI: \textit{``The big danger to our students is that they just sit in the basement and don't interact anymore with other students or faculty, instead developing relationships with an AI...it could be toxic, wear down the psychological health of our students.''} Other participants were ambivalent, noting, \textit{``I just need a general persona or tone. I don't need to make it more academic or professional. It's not important'' (P3).}

\subsection{Where instructors diverged on priorities}

While instructors generally desired the embedding of custom course materials and pedagogical strategies, the card-sorting of categories like \textbf{personalization}, \textbf{course management}, \textbf{task/objective}, and \textbf{user environment} produced highly mixed rankings. Many instructors explicitly noted that their courses were ``weird'' (P5) or ''different'' (P1, P9, P10), or that they preferred employing tailored pedagogical methods that required calibration to their classroom and material (P6, P7, P10). P9 described some of her courses as ``open-world,'' while others were structured. These divergences reflected differences in course size, format, and teaching style, indicating that customization needs for chatbots are not uniform.

\subsubsection{Instructors were polarized on \textbf{personalization}.}

Instructors often differed vastly on prioritizing the \textbf{personalization} category for a variety of reasons, from class format to course size to personal preference. P10, who teaches a large-scale Calculus course, considered personalization a high-priority for keeping students engaged, which she considered a big challenge: \textit{``Personalization for students is part of motivating students to attend real class, encouraging participation.''} P6, who teaches a large-scale data science course, gave an example of how he would want the chatbot to be personalized to variation in students' pre-requisite knowledge, \textit{``My ideal chatbot would be like [to the student], `You haven't learned geospatial D3 function calls. Here's a brief primer on the top three.'''} 

Some participants had a more nuanced view. P9, who teaches a project-based interaction design course, struggled to rank \textbf{personalization}, swapping it between high and medium priority multiple times. She ultimately decided it depends on course format, explaining that some of her courses are \textit{``open-world, more free-range. Some students come in as complete experts and some just want to be in the room.''} For those courses, she prioritized personalizing content for students because students come from a range of backgrounds and have different goals. She added that prioritizing \textbf{personalization} in custom chatbots would also differ \textit{``if something was asynchronous versus in-person. Personalization is lower for lecture-based formats compared to asynchronous because having lectures all be the same is really important, so students can talk about it afterwards. They need some kind of constant.''} 

However, unlike P9 and P10, P5 did not think personalization would be useful in his project-based course: \textit{``My class is weird. This would be great for a math class...but mine not as much.''} Other instructors shared that personalization was not critical, or even unhelpful, in scaled-up course settings. P1, who has taught computing courses of multiple sizes, explained, \textit{``It's difficult when I think of it in the scaled classroom of 600 students. I'm wary of the personalization aspect. Sometimes, challenges are good for students when it's not personalized, they need to stumble upon some things and make their way through it. It's just a part of their learning experience.''} One participant felt strongly against personalization, sharing that the ``depersonalization'' of these chatbots was of greater importance to him (P2). He reiterated the value of human-to-human engagement in educational settings over student-personalized chatbots: \textit{``We need to be a community of humans that learn together.''}

\subsubsection{Instructors had divergent needs regarding customizing \textbf{course management}, \textbf{user environment/platform}, and \textbf{task/objective}.}

Some instructors found \textbf{course management} more important than others, and course size and format were key factors in their prioritization. For example, P6, who taught a large data science course, described the challenge of \textit{``making sure I'm not the bottleneck for releasing assignments, grades, or logistic questions...minimizing myself as the single blocker in the class.''} Other instructors believed that \textbf{course-management} customization should free up capacity for \textit{``maximizing one-on-one time with students''} (P8, P9). Instructors of large lecture courses valued the customization of management functions: dealing with regrade requests (P10), automatically answering logistical questions on class forums (P6, P8, P10) or managing student groups (P4, P8, P10). Instructors shared how coordinating large classes could be ``nightmarish'' (P10), with overwhelming emails, logistical issues, and questions --- which they hoped chatbots could help alleviate. Instructors of smaller or project-based courses ranked \textbf{course management} lower, prioritizing other needs more; they explained that they could handle logistics themselves or with a TA rather than delegate to a chatbot, although they thought that these features could still be useful. For example, helping students set up course software (P5) or collecting metadata (P1). Instructors who taught more than one course explicitly observed that they would prioritize this feature differently depending on course size: \textit{``It could be a big help in saving time on these tasks, like answering individual questions once there are a couple hundred students'' (P4).}

In terms of adapting chatbots to specific user environments (e.g. Canvas, Discord, Piazza, Moodle, in-person labs), some instructors found context-sensitive customization was essential for adoption: \textit{``If all of your students are already on Discord, I would want [the chatbot] to customize to wherever the students are for the class'' (P8).} On the other hand, others ranked it low, suggesting functionality mattered more than where the chatbot was deployed: \textit{``Even if students had to go to a separate [University]GPT.com to use the chatbot, I'd be okay with that.''} 

Instructors also diverged on the importance of aligning chatbots with specific learning \textbf{tasks/objectives}. Some saw explicit task alignment as critical for keeping interactions on track, instead of leaving the chatbot open-ended (P1, P4). However, others ranked it low, reasoning that if course materials and pedagogical strategies were embedded effectively, the chatbot would naturally support the right objectives without requiring additional configuration.

\subsubsection{Additional categories proposed by instructors.} 

Participants proposed five additional categories or features. P10, who teaches a foundational mathematics course to thousands of students, described her role as more of a course coordinator than instructor, and specifically wanted to customize the chatbot towards assisting instructional training: \textit{``A lot of effort goes to teaching instructors who might not share similar training backgrounds, who don't trust flipped classrooms or active learning. If the chatbot can help me explain to instructors, it would be helpful.''} Other instructors like P1 emphasized cross-disciplinary relevancy, noting that she taught a CS course for non-CS students and wanted to help them map computing concepts to their own fields. Instructors also brought up cost and budget of relying on such a chatbot (P10), adaptability to student intent (P2), and easy set-up (P2).

These divergences illustrate that while some instructor needs converged, other major priorities were highly context-dependent. This reinforces that chatbot customization cannot follow a one-size-fits-all model, and that design tools must balance support for common functions with flexibility for diverse teaching contexts.

\subsubsection{Instructors wanted to share configurations of chatbots with colleagues but prioritization varied.}

The ability to share and open-source custom configurations (\textbf{community sharing}) was seen as important but unevenly prioritized. Every instructor expressed willingness to share their chatbot configurations, with no one citing personal or intellectual restrictions. Some ranked shareability as a high priority, framing it as a way to reduce duplicated effort and accelerate course design, while others placed it lower to prioritize other immediate needs. As P5 explained, \textit{``Sharing is caring. Instructors have their own ideas of how to do stuff anyways...There's nothing in here I want private and for myself.''} P2 shared this sentiment and framed it as he would for regular course materials: \textit{``It's the 3 R's right. Reuse, remix, repurpose.''} Instructors noted an important exception: student work should always remain private (P6, P9, P10). 

Concerns or uncertainty brought up by participants were largely logistical. Participants desired to share configurations of their chatbot because there is no current systematic, efficient way to browse, adapt, or remix colleagues’ materials, even when they wanted to learn from others. P8 noted, \textit{``Most are very happy to share their materials, but adoption is hard. Like, so-and-so is willing to give me their materials, but now I have to integrate that into my class.''} 









\section{Discussion}

Our findings revealed convergence on some priorities (e.g., course materials, pedagogy, guardrails) but broad divergence on others (e.g., personalization, course management), showcasing the heterogeneity of instructional contexts. This range of needs directly reflects the constraints and preferences that instructors must generally contend with when it comes to running their own courses. One instructor (P9) labeled her course ``open-world,'' a descriptor that can also apply to instruction. There is no ``one way'' to teach, which parallels prior findings of instructors desiring chatbot adaptability~\cite{hedderich2024a}. Our findings also suggest that, although reusable chatbot components are feasible for common needs, modularity and flexibility are essential to accommodate divergent requirements. There emerges a paradox of customization; building highly tailored technologies makes them difficult to scale or adapt across end-users (instructors). Participants desired easy configuration, echoing prior findings~\cite{vermette2019freedom}, as too many customization options can overwhelm already overloaded instructors. This tension has been explored in HCI around configurability and end-user programming~\cite{ko2004six}. Designing classroom chatbots thus necessitates balance between reusability and flexibility to reduce adoption costs. In the following sections, we discuss why instructor needs struggle to be met by existing AI systems, as well as forward-looking design implications.

\subsection{Why instructor needs struggle to be easily met by existing AI systems}

Instruction is dynamic. It naturally evolves with new material, updated pedagogical techniques, changing cohorts of students, and instructors' pedagogical choices. Several instructors noted that their courses were ``weird'' (P5) or ``different'' (P1, P9, P10), requiring adaptation of existing resources for their own needs. Like prior studies~\cite{vermette2022uncovering}, they wanted to ``reuse, remix, and repurpose'' (P2). The \textit{ephemerality} of instruction is thus difficult to capture in fixed chatbot systems. Moreover, instruction is characterized by \textit{multiplicity}; even within the same course, instructors may adopt different pedagogical strategies, logistical practices, or disciplinary framings. 

Existing AI chatbots are not designed to accommodate these factors. While AI course assistants are highly valuable, they have focused on providing student pedagogical support~\cite{labadze2023role, kazemitabaar2024codeaid, atmosukarto2021enhancing}. Our findings reveal a broader set of instructor needs that includes balancing logistics and learning, supporting collaboration across instructors, and adapting to the evolving nature of teaching itself. While existing AI tools are well-suited to contexts where course content and practices are relatively stable, they are less equipped to evolve with this dynamic, heterogeneous environment. 

\rev{Another challenge instructors face is that existing AI systems are often ``all-or-nothing;'' instructors must choose between investing effort in a bespoke chatbot or relying on generic public models. The convergence and divergence we observed illustrate why this binary approach is insufficient. Convergent needs (e.g. embedding course materials, configuring pedagogical strategies, establishing guardrails) represent chatbot capabilities that many instructors expect and could be templatized into reusable building blocks. Divergent needs, however, reveal contextual priorities that vary greatly across instructors, such as when personalization is valuable, or whether platform integration is needed. These patterns also show that categories do not transfer cleanly onto one-to-one agent roles; instructors sometimes described multiple forms of customization within a single teaching task (e.g. P6's ideal chatbot combines both pedagogical strategies + personalization). Together, this indicates a structural mismatch between current chatbot designs and the ways instructors desire customization capabilities across instructional contexts.}

\subsection{Design implications towards teachable teammates}

Instructors seek to customize AI chatbots, but they need feasible, low-barrier, and sustainable ways to do so. Technical building blocks already exist --- approaches such as retrieval-augmented generation (RAG)~\cite{lewis2020retrieval} and multi-agent orchestration frameworks~\cite{wu2024autogen} demonstrate how knowledge access can be modularized and agent roles distributed. In the introduction, we described our vision of \textit{teachable teammates}, or modular agents that can be recruited into a roster and bound to course rules, roles, and materials (e.g. grader bot bound to rubrics, lab coach bot with safety protocols). Our design implications show how instructor-articulated needs map onto this vision, extending prior technical work by grounding modularity in the realities of instruction. 

\begin{itemize}
    \item 1) \textbf{Modularize customization into re-usable building blocks.} Instructors converged on core needs like \textbf{constraints and guardrails}, \textbf{pedagogical strategy}, and \textbf{course-specific material}. Modularizing these into configurable blocks could reduce redundant effort and make customizations easier to remix within communities of practice, while preserving room for course idiosyncrasies.  
    \item 2) \textbf{Structure chatbots as agentic workflows instead of monolithic systems.} Prompt-based customization often ties stable requirements (e.g., grading policies) to changing course content, requiring unnecessary rework. Agentic workflows can decouple these layers, making customization more sustainable and supporting easier iteration, testing, and long-term maintenance for busy instructors.  
    \item 3) \textbf{Support instructor sharing while accommodating other important stakeholders.} Many instructors desired to share configurations to foster reuse and remixing but few brought up institutional barriers. Given that instruction involves multiple stakeholders (e.g. institutions, students), design for open-source or peer-to-peer sharing must also confront institutional constraints and responsibilities around privacy, accountability, and ownership. 
\end{itemize}

\section{Conclusion}

Through interviews and a card-sorting activity, we explored how instructors want to customize AI chatbots, finding convergence around guardrails, pedagogical strategies, and course materials. Other areas of desired customization (e.g. personalization, course management) were diverse, indicating that there is no one-size-fits-all chatbot design. Our findings point to design opportunities around modular building blocks, agentic workflows, and mechanisms for sharing. These directions suggest that chatbots should be framed less as monolithic systems and more as \textit{teachable teammates} that can be recruited and adapted to different pedagogical contexts. Future tools building on this vision may lead to classroom chatbots that are more sustainable, adaptable, and aligned with the creativity and agency of instructors.

\bibliographystyle{ACM-Reference-Format}
\bibliography{ref}

\end{document}